\documentclass[prd,nofootinbib,12pt]{revtex4}

\usepackage{lineno,hyperref}
\usepackage{amsfonts,amssymb,amsmath}
\usepackage{epsfig}
\usepackage{mathrsfs}
\usepackage{amssymb}
\usepackage{graphicx}
\usepackage{amsmath}
\usepackage{graphicx}
\usepackage{amsmath}
\usepackage{xcolor}
\usepackage{color}
\usepackage{appendix}

\begin{document}

\title{Krylov complexity of thermal state in early universe}

\author{Tao Li$^{1}$}
\author{Lei-Hua Liu$^{1}$}
\email{liuleihua8899@hotmail.com}
\affiliation{$^1$Department of Physics, College of Physics, Mechanical and Electrical Engineering, Jishou University, Jishou 416000, China}

\begin{abstract}
	
Thermal interactions are ubiquitous in the cosmos, driving systems toward equilibrium. In this work, we investigate the evolution of thermal states across the early universe, encompassing the inflationary, radiation-dominated (RD), and matter-dominated (MD) eras, through the lens of Krylov complexity. Utilizing a purification scheme, we map the thermal state to a two-mode pure state, facilitating an open-system analysis of Krylov complexity in contrast to closed-system methodologies. Our numerical results demonstrate that Krylov complexity grows exponentially during inflation, indicating chaotic behavior, before saturating at nearly constant values in the RD and MD eras due to particle production via preheating. Furthermore, we analyze the Krylov entropy, which exhibits an evolutionary trend analogous to that of complexity. Crucially, our analysis reveals a dynamical transition in the universe's dissipative nature: with the universe acting as a strongly dissipative system during inflation and transitioning to a weakly dissipative regime in the subsequent eras. These findings provide a novel quantum information perspective on early universe dynamics.

\end{abstract}

\maketitle


\section{introduction}
\label{introduction}

With the rapid advancement of quantum complexity theory in high-energy physics, several distinct approaches to quantifying complexity have emerged. The first prominent framework is \textit{quantum circuit complexity}~\cite{Susskind:2014moa,Jefferson:2017sdb,Nielsen:2005mkt,Aaronson:2016vto}, defined as the minimal number of elementary quantum gates required to synthesize a target state from a reference state. Nielsen and collaborators pioneered a geometric approach to this problem~\cite{Nielsen:2005mkt,Nielsen:2006cea,Dowling:2006tnk}, reformulating complexity as an optimization problem over continuous paths in Hilbert space. A complementary approach is provided by \textit{Krylov complexity} (or K-complexity)~\cite{Parker:2018yvk,Barbon:2019wsy,Avdoshkin:2019trj,Balasubramanian:2022tpr}, which quantifies the complexity of operator evolution under Heisenberg dynamics. Unlike circuit complexity, Krylov complexity characterizes the growth of operators within the Krylov subspace~\cite{Parker:2018yvk}, generated through iterative applications of the Liouvillian operator.

Closely related to this framework is the concept of \textit{Krylov entropy}, which serves as a potent measure of quantum chaos by quantifying the delocalization of an operator within its Krylov basis~\cite{Balasubramanian:2022tpr,Rabinovici:2023yex}. This delocalization directly probes information scrambling and chaotic dynamics. Notably, these approaches are fundamentally interconnected: recent work demonstrates that spread complexity (equivalent to Krylov complexity) emerges as a specific limit of circuit complexity, establishing a unified framework built from the fundamental operations of time-evolution and superposition~\cite{Beetar:2025mdz}.

In recent years, Krylov complexity has witnessed rapid development across both condensed matter and high-energy physics. In Ref.~\cite{Muck:2022xfc}, a systematic study was conducted on Krylov complexity across various polynomial bases constructed via the Lanczos algorithm. Applications of K-complexity span a wide range of theoretical landscapes, including the SYK model~\cite{Rabinovici:2020ryf,Jian:2020qpp,He:2022ryk}, generalized coherent states~\cite{Patramanis:2021lkx}, spin chains such as the Ising and Heisenberg models~\cite{Cao:2020zls,Trigueros:2021rwj,Heveling:2022hth}, conformal field theory~\cite{Dymarsky:2021bjq,Caputa:2021ori}, and topological phases of matter~\cite{Caputa:2022eye}. Chaos has been interpreted within this framework as a form of delocalization in Krylov space~\cite{Dymarsky:2019elm}. While Ref.~\cite{Bhattacharjee:2022vlt} demonstrated that exponential growth of Krylov complexity can occur in integrable systems exhibiting saddle-dominated scrambling, a behavior corroborated for spread complexity in Ref.~\cite{Huh:2023jxt}, recent work suggests that this behavior may break down at late times~\cite{Aguilar-Gutierrez:2025hbf}. Further implications and extensions of Krylov complexity have been explored in numerous recent studies~\cite{Erdmenger:2023wjg,Zhai:2024odw,Hashimoto:2023swv,Vasli:2023syq,Gill:2023umm,Bhattacharjee:2023uwx,Adhikari:2022whf,Loc:2024oen,Caputa:2024vrn,Basu:2024tgg,Sasaki:2024puk,Caputa:2024xkp,Bhattacharjee:2022qjw,Sahu:2024opm,Kim:2021okd,Chen:2024imd,Bhattacharjee:2024yxj,Sanchez-Garrido:2024pcy,Chattopadhyay:2024pdj,Camargo:2023eev,Camargo:2024deu,Baggioli:2024wbz,Alishahiha:2025sep,Heller:2025ddj,Grabarits:2025xys,Chhetriya:2025ndi,Liu:2025caj,He:2025guu,Angelinos:2025drf,Demulder:2025uda,FarajiAstaneh:2025thi,Beetar:2025erl,Imani:2025etp,Heller:2024ldz}. For a comprehensive overview, we refer readers to recent review articles~\cite{Rabinovici:2025otw,Nandy:2024htc}.

This paper aims to investigate Krylov complexity in the context of the early universe, encompassing the inflationary period, the RD era, and the MD era. While there has been significant research addressing circuit complexity in cosmology~\cite{Choudhury:2020hil, Bhargava:2020fhl, Lehners:2020pem, Bhattacharyya:2020rpy, Adhikari:2021ked}, showing, for instance, oscillations during inflation and similar post-inflationary trends~\cite{Liu:2021nzx, Li:2021kfq, Li:2023ekd}, as well as irregular oscillations when quantum gravitational effects are included~\cite{Li:2023ekd}, studies on Krylov complexity in this domain remain sparse. Ref.~\cite{Adhikari:2022oxr} investigated K-complexity with varying speeds of sound during inflation using a \textit{closed-system} methodology. However, treating the universe strictly as a closed system presents theoretical challenges. The ADM energy is well-defined and conserved only in asymptotically flat spacetimes; it remains undefined for cosmological spacetimes lacking suitable asymptotic boundaries. While the energy-momentum tensor is covariantly conserved given time-translation symmetry~\cite{Nother:1933mma}, the explicit time-dependence of the metric in cosmology leads to the global non-conservation of the integrated energy component ($T_{00}$)~\cite{Cheung:2007st}. Consequently, an \textit{open-system} methodology appears more physically robust than a closed-system approach. 

In this work, we treat quantum phenomena during inflation under the local approximation of a non-gravitating universe~\cite{Chakraborty:2025izq}, an approach that has motivated recent applications of open quantum systems to inflation and gravity~\cite{Salcedo:2024smn,Salcedo:2025ezu}. Utilizing this open-system framework, Ref.~\cite{Li:2024kfm} previously examined Krylov complexity with modified dispersion relations during inflation, identifying the inflationary period as a strongly dissipative system where complexity generally increases with the scale factor. A key aspect of our current study is the construction of the wave function using the open-system methodology outlined in Ref.~\cite{Bhattacharya:2022gbz,Li:2024kfm,Zhai:2024tkz,Zhai:2025abc,Liu:2025caj}, which serves as our starting point for analyzing thermal states.

In the early universe, high temperatures drove cosmic expansion and particle production, making the incorporation of thermal effects into the wave function both essential and well-motivated. Ref.~\cite{Kar:2021nbm} formally introduced the notion of thermal Krylov complexity. In a cosmological context, Ref.~\cite{Wang:2022mix} found that circuit complexity decreases during the early universe. Building on this, we previously investigated the Krylov complexity of a thermal state during inflation~\cite{Li:2024iji}. While Krylov complexity can be extended to quantum field theory for a scalar field at arbitrary temperatures~\cite{He:2024xjp}, the analysis in Ref.~\cite{Li:2024iji} was confined strictly to the inflationary epoch. To capture the full dynamical history, we extend the Krylov complexity analysis here to include the evolution through the RD and MD eras. Furthermore, as noted in Ref.~\cite{Chapman:2024pdw}, the Lyapunov exponent can exhibit non-monotonic behavior, potentially rendering it an unreliable sole measure of chaos. In this work, we therefore anticipate that the open-system nature of the expanding universe will imprint a distinctive, non-monotonic profile on the Lyapunov exponent.

The paper is organized as follows. In Sec.~\ref{sec: Krylov complexity and Lanczos algorithm with an approach of closed system}, we introduce the Lanczos algorithm within the framework of a closed system. A brief overview of the early universe, including the inflationary, RD, and MD eras, along with the corresponding scale factor evolution, is provided in Sec.~\ref{sec:Single-Component Universe}. The thermal state formalism is discussed in Sec.~\ref{sec:The wave function of thermal state}. We then investigate Krylov complexity using the closed-system approach in Sec.~\ref{sec: Krylov complexity with an approach of closed system}. The Lanczos coefficients and Krylov complexity obtained via the open-system methodology are presented in Sec.~\ref{sec:Krylov complexity with the approach of open system}. Finally, Sec.~\ref{Summary and outlook} provides concluding remarks and an outlook. Throughout this work, we adopt Planck units where $c=G=k_B=1$.


\section{ Lanczos algorithm with closed-system methodology}
\label{sec: Krylov complexity and Lanczos algorithm with an approach of closed system}

This section formulates the Lanczos algorithm within the framework of a closed quantum system. We begin by expressing the time evolution of an operator $\mathcal{O}(t)$ in the Heisenberg picture:
\begin{equation}
\partial _{t}\mathcal{O}(t)=i[H,\mathcal{O}(t)],
\label{Ot}
\end{equation}
with the formal solution given by
\begin{equation}
\mathcal{O}(t)=e^{iHt}\mathcal{O}e^{-iHt}.
\end{equation}
We define the Liouvillian superoperator $\mathcal{L}$ acting on an operator $Y$ as $\mathcal{L}_Y \equiv [H, Y]$, which shares the same commutator structure as Eq.~\eqref{Ot}. Using this definition, the time-evolved operator $\mathcal{O}(t)$ can be expanded via the Baker-Campbell-Hausdorff formula:
\begin{equation}
\mathcal{O}(t) = e^{i\mathcal{L}t}\mathcal{O} = \sum_{n=0}^{\infty } \frac{(it)^{n}}{n!} \mathcal{L}^{n}\mathcal{O}(0) = \sum_{n=0}^{\infty } \frac{(it)^{n}}{n!} |\tilde{\mathcal{O}}_{n}),
\end{equation}
where we have introduced the notation $|\tilde{\mathcal{O}}_n) \equiv \mathcal{L}^n\mathcal{O}$ to represent the basis states. These states form the Krylov subspace:
\begin{equation}
\mathcal{O}\equiv |\tilde{\mathcal{O}}_{0}) ,\quad \mathcal{L}^{1}\mathcal{O}\equiv |\tilde{\mathcal{O}}_{1}) ,\quad \mathcal{L}^{2}\mathcal{O}\equiv |\tilde{\mathcal{O}}_{2}),\quad \dots \quad .
\end{equation}
Since this naive basis is generally non-orthogonal, we employ the Lanczos algorithm (a specific instance of the Gram--Schmidt procedure) to construct an orthonormal basis $\{|\mathcal{O}_n)\}$, referred to as the Krylov basis~\cite{Parker:2018yvk}. The process begins with:
\begin{equation}
|\mathcal{O}_{0}) = |\tilde{\mathcal{O}}_{0}) = \mathcal{O}, \quad |\mathcal{O}_{1}) = b_{1}^{-1}\mathcal{L}|\mathcal{O}_{0}),
\end{equation}
where $b_{1} = \sqrt{(\mathcal{L}\mathcal{O}_{0}|\mathcal{L}\mathcal{O}_{0})}$ is the normalization constant. Subsequent basis vectors are generated via the recurrence relation:
\begin{equation}
|A_{n}) = \mathcal{L}|\mathcal{O}_{n-1}) - b_{n-1} |\mathcal{O}_{n-2}),
\label{iterative relation}
\end{equation}
with the normalization condition and Lanczos coefficients defined as:
\begin{equation}
|\mathcal{O}_{n}) = b_{n}^{-1} |A_{n}), \quad b_{n} = \sqrt{(A_{n}|A_{n})}.
\end{equation}
The iteration in Eq.~\eqref{iterative relation} halts if $b_n = 0$, yielding a finite-dimensional Krylov subspace. The operator $\mathcal{O}(t)$ can then be expanded in this orthonormal basis as:
\begin{equation}
\mathcal{O}(t) = e^{i\mathcal{L}t}\mathcal{O} = \sum_{n=0}^{\infty } i^{n} \phi_{n}(t)|\mathcal{O}_{n}),
\end{equation}
where $\phi_{n}(t)$ represents the real transition amplitude, serving as the ``wave function'' in the Krylov space, subject to the normalization $\sum_{n}|\phi_{n}|^{2}=1$. Substituting this expansion into the Heisenberg equation~\eqref{Ot}, we obtain the discrete Schr{\"o}dinger equation:
\begin{equation}
\partial_{t}\phi_{n}(t) = b_{n}\phi_{n-1} - b_{n-1}\phi_{n+1}.
\end{equation}
Based on this probability distribution, the Krylov complexity is defined as the average position on the Krylov chain:
\begin{equation}\label{eq:3.11}
K \equiv \sum_{n} n |\phi_{n}|^{2}. 
\end{equation}
According to the hypothesis in Ref.~\cite{Parker:2018yvk}, for chaotic systems, the Lanczos coefficients are linearly bounded asymptotically:
\begin{equation}
b_{n} \le \alpha n + \gamma,
\end{equation}
where $\alpha$ is the growth rate and $\gamma$ is a constant intercept. The system's chaotic behavior is quantified by the Lyapunov exponent $\lambda_L$, which is related to the growth rate $\alpha$ by:
\begin{equation}
\lambda_L = 2\alpha.
\label{gamma}
\end{equation}
Before proceeding to calculate the Krylov complexity, it is necessary to specify the evolution of the scale factor across the distinct cosmological eras.

\section{FUNDAMENTALS OF COSMOLOGICAL EVOLUTION}
\label{sec:Single-Component Universe}
In this section, we review the fundamental dynamics of the early universe, adopting the framework outlined in Ref.~\cite{Baumann:2009ds}. We specifically examine the successive epochs of inflation, RD, and MD, which are characterized by the evolution of the scale factor $a(\eta)$ with respect to conformal time $\eta$.

We begin with the Friedmann-Lema\^{i}tre-Robertson-Walker (FLRW) metric, which provides an accurate description of the geometry for a spatially homogeneous and isotropic universe:
\begin{equation}
ds^{2} = a(\eta)^{2}(-d\eta^{2}+d\vec{x}^{2}).
\label{FRW metric1}
\end{equation}
Within this metric framework, we analyze the Krylov complexity across distinct cosmological epochs. Following Ref.~\cite{Baumann:2009ds}, we model the evolution as a sequence of distinct eras, each dominated by a single fluid component (sequentially: inflation, RD, and MD). The equation of state parameter is defined as $w_I = P_I / \rho_I$, where the subscript $I$ denotes inflation, RD, or MD. To establish the explicit relationship between the scale factor and conformal time, we consider the comoving particle horizon, defined as:
\begin{equation}
\label{comoving distance}
 \chi_{ph}(\eta) \equiv \int_{0}^{\eta} d\eta' = \int_{\ln{a_{i}}}^{\ln{a}}(aH)^{-1}d\ln{a},
\end{equation}
where the integrand is governed by the comoving Hubble radius $(aH)^{-1}$. Assuming a constant equation of state $w$, the comoving Hubble radius evolves as $(aH)^{-1} = H_{0}^{-1}a^{\frac{1}{2}(1+3w)}$ (where we drop the subscript $I$ for generality). Integrating Eq.~\eqref{comoving distance}, we obtain the dependence of conformal time on the scale factor:
\begin{equation}
\eta = \frac{2H_{0}^{-1}}{(1+3w)} a^{\frac{1}{2}(1+3w)}.
\label{explicit relation}
\end{equation}
Consequently, we derive the explicit solutions for the distinct epochs:
\begin{equation}
\begin{split}
\eta = \left\{\begin{matrix}
& aH_{0}^{-1} \quad (w =\frac{1}{3}), & \mathrm{RD} \\
& 2\sqrt{a}H_{0}^{-1} \quad (w=0), & \mathrm{MD}
\end{matrix}\right.
\end{split}
\label{eq:conformal time and scale factor}
\end{equation}
For the inflationary epoch, the relation simplifies to $\eta \approx -(aH_0)^{-1}$ (for $w \approx -1$). Utilizing these three piecewise solutions, we can evaluate the Krylov complexity across the entire early universe, applying matching conditions at the transition boundaries. Further details regarding the dynamics of these eras are provided in Appendix \ref{sec: inflation, RD and MD}.

\section{Wave function of the thermal state }
\label{sec:The wave function of thermal state}

We investigate the Krylov complexity of the thermal state, defined as:
\begin{equation}
\hat{\rho} = \frac{1}{Z(\beta)} \sum^{\infty}_{n=0} e^{-\beta E_{n}} |n\rangle \langle n|,
\label{eq:Density_equation}
\end{equation}
where $Z(\beta) = \text{Tr}(e^{-\beta \hat{H}})$ denotes the partition function. In the context of bosonic oscillators, where $\beta = 1/( T)$ and the energy levels are given by $E_n = n \omega$ (with $\omega$ representing the frequency), we adopt the approach utilized in Refs.~\cite{Jefferson:2017sdb, Ali:2018fcz, Haque:2021kdm} to treat the thermal state within a pure-state formalism. Since $\hat{\rho}$ is inherently a mixed state, we employ the purification method to construct a corresponding pure state, denoted as $|\Psi\rangle$. This is achieved by extending the original Hilbert space to a doubled space $\mathcal{H}_{\text{total}} = \mathcal{H} \otimes \mathcal{H}_{\text{anc}}$, where $\mathcal{H}_{\text{anc}}$ represents the ancillary degrees of freedom. The physical mixed state $\hat{\rho}$ is recovered by taking the partial trace of $|\Psi\rangle$ over the ancillary space $\mathcal{H}_{\text{anc}}$, confirming that $|\Psi\rangle$ serves as a valid purification of the thermal state in Eq.~\eqref{eq:Density_equation}. A similar methodology is discussed in Ref.~\cite{Das:2024zuu}.

It is crucial to emphasize that the expectation value of any system observable $\hat{\mathcal{O}}$ remains invariant under purification. This consistency is guaranteed by the relation
\begin{equation}
\langle \hat{\mathcal{O}} \rangle \equiv \operatorname{Tr} \big( \hat{\rho} \hat{\mathcal{O}} \big) = \langle \Psi | \hat{\mathcal{O}} \otimes \mathbb{I}_{\text{anc}} | \Psi \rangle,
\end{equation}
which ensures the reliability of computing Krylov complexity within this pure-state framework. However, the purification $|\Psi\rangle$ is not unique due to the freedom in choosing the ancillary Hilbert space $\mathcal{H}_{\text{anc}}$. In this work, we adopt the standard choice of duplicating the original Hilbert space, i.e., $\mathcal{H}_{\text{anc}} = \mathcal{H}$. This leads directly to the standard thermofield double (TFD) state:
\begin{equation}
 | \text{TFD} \rangle \equiv \frac{1}{Z}\sum^{\infty}_{n=0} e^{-\beta E_{n}/2} |n \rangle \otimes |n \rangle_{\text{anc}}. 
\label{eq:Mixed_state}
\end{equation}
To align this state with the generalized definition of a two-mode squeezed vacuum, we introduce specific phase factors into the superposition. The generalized purification of the thermal state can then be expressed as:
\begin{equation}
| \Psi \rangle_{\phi} = \frac{1}{Z}\sum^{\infty}_{n=0}(-1)^{n}e^{-2in\phi_{k}}e^{-\beta E_{n}/2} |n \rangle \otimes |n \rangle_{\text{anc}},
\label{eq:generalized TFD}
\end{equation}
where the term $(-1)^n e^{-2in\phi_k}$ represents the relative phase factors. This state can be identified as a two-mode squeezed vacuum state generated by the squeezing operator $\hat{S}(\xi_k)$:
\begin{equation}
| \Psi \rangle_{\phi} = \hat{S}(\xi_k) |0 \rangle \otimes |0 \rangle_{\text{anc}} = \frac{1}{\cosh{r_{k}}}\sum^{\infty}_{n=0}(-1)^{n}e^{-2in\phi_{k}}\tanh^{n}{r_{k}} |n \rangle \otimes |n \rangle_{\text{anc}}.
\label{eq:two_mode_squeezed_state}
\end{equation}
Comparing the coefficients in Eq.~\eqref{eq:two_mode_squeezed_state} and Eq.~\eqref{eq:generalized TFD}, we derive the correspondence between the squeezing parameter $r_k$ and the effective temperature $T = 1/\beta$:
\begin{equation}
\beta\omega = -\ln{\left(\tanh^{2}{r_{k}}\right)}, \quad \cosh r_{k} = \frac{1}{\sqrt{1-e^{-\omega/T}}},
\label{corresponding relation}
\end{equation}
where $\omega$ denotes the frequency. Consequently, substituting these relations back, we obtain the explicit wave function:
\begin{equation}
| \Psi \rangle_{\phi} = \sqrt{1-e^{-\frac{\omega}{T}}}\sum^{\infty}_{n=0}e^{-2in\phi_{k}}e^{-\frac{n\omega}{2T}} |n \rangle \otimes |n \rangle_{\text{anc}}.
\label{eq:wave function of close syetem}
\end{equation}
This pure state wave function serves as the foundational framework for investigating Krylov complexity and K-entropy. Specifically, we define the Krylov basis states using the operator state representation $|\mathcal{O}_n\rangle \sim |n\rangle \otimes |n\rangle_{\text{anc}}$. Following the open-system methodology developed in Ref.~\cite{Bhattacharya:2022gbz}, this construction will be further applied to systematically derive the wave function evolution.

 \section{Evolution of effective temperature and squeezed angle}
 \label{sec:Evolution of temperature and squeezed angle}

 In this section, we analyze the dynamics of the squeezing parameters, specifically $r_k$ and $\phi_k$, following conventional approaches. However, our scenario presents a distinctive feature: the parameter $r_k$ is intrinsically determined by the effective temperature and frequency through the wave function derived in Eq.~\eqref{eq:wave function of close syetem}. Notably, the evolution of $r_k$ can be effectively mapped onto the effective temperature dynamics. This mapping arises from the fundamental relation:
 \begin{equation}
 \beta\omega = -\ln(\tanh^2 r_k),
 \end{equation}
 which allows the frequency dependence to be absorbed into the effective temperature parameter.
 
 Prior to analyzing the effective temperature dependence and phase evolution, we first examine the Krylov complexity within the framework of single-field inflation. Our analysis begins with the metric formulation presented in Eq.~\eqref{FRW metric1}. In contrast to previous approaches such as Ref.~\cite{Li:2024kfm}, we deliberately avoid employing the full Mukhanov-Sasaki variable. This choice is motivated by the need to isolate the explicit influence of the inflationary potential on Krylov complexity, which is often obscured in the Mukhanov-Sasaki formalism due to the mixing of metric and scalar perturbations.
 
 To clarify this distinction, recall that the standard Mukhanov-Sasaki variable is defined as:
 \begin{equation}
  v_{\text{MS}} \equiv a\left( \delta\phi + \frac{\dot{\phi}}{H} \Psi \right), 
 \end{equation}
 where $\phi$ is the inflaton field, $\Psi$ is the metric perturbation in the Newtonian gauge, and $H$ is the Hubble parameter. The variable $v_{\text{MS}}$ implicitly contains contributions from the inflationary potential through the background equation terms $\dot{\phi}/H$, where the background dynamics satisfy:
 \begin{equation}
 H^2 = \frac{1}{3M_{\text{Pl}}^2}\left(\frac{1}{2}\dot{\phi}^2 + V(\phi)\right), \quad \ddot{\phi} + 3H\dot{\phi} + V_{,\phi} = 0.
 \end{equation}
 
 To maintain direct visibility of the potential's role, we adopt the **spatially flat gauge** (where metric perturbations are neglected, $\Psi \approx 0$) and consider the rescaled perturbation of the inflaton field directly:
 \begin{equation}
 \phi(\eta,x) = \bar{\phi}(\eta) + \frac{v(\eta,x)}{a(\eta)},
 \label{eq: perturbation of inflaton}
 \end{equation}
 where $\bar{\phi}(\eta)$ represents the homogeneous background field and $v(\eta,x)$ denotes the field perturbation (distinct from the full gauge-invariant variable $v_{\text{MS}}$). This approach ensures that the specific contributions of the inflationary potential to Krylov complexity remain analytically transparent.
 
 The action for the inflaton field is given by:
 \begin{equation}
 S = \int d\eta d^3 x \sqrt{-g} \bigg[ \frac{1}{2}g^{\mu\nu}\partial_\mu\phi\partial_\nu\phi - V(\phi) \bigg],
 \label{eq: total action}
 \end{equation}
 where $g^{\mu\nu}$ is the metric corresponding to Eq.~\eqref{FRW metric1}. Substituting the perturbation ansatz from Eq.~\eqref{eq: perturbation of inflaton} into the action Eq.~\eqref{eq: total action} and expanding to second order, we obtain the following quadratic action after performing integration by parts:
 \begin{equation}
 S^{(2)} = \frac{1}{2}\int d\eta d^{3}x \left[ v'^2 - (\partial_i v)^2 + \left( \frac{a''}{a} - a^2 V_{,\phi\phi} \right) v^2 \right],
 \label{eq: second order action}
 \end{equation}
 where primes denote derivatives with respect to conformal time $\eta$, and $V_{,\phi\phi} \equiv d^2V/d\phi^2$ represents the second derivative of the potential. A crucial advantage of this formulation is the explicit separation of the geometric mass term ($a''/a$) and the potential-dependent mass term ($a^2 V_{,\phi\phi}$).
 
 \begin{itemize}
 	\item \textbf{Inflationary Era}: The term $a^2 V_{,\phi\phi}$ is typically negligible compared to the geometric term during inflation, consistent with the slow-roll approximation.
 	\item \textbf{Radiation and Matter Dominated Eras}: This potential term becomes significant during the RD and MD eras, where slow-roll approximations break down, and the potential contribution must be rigorously accounted for.
 \end{itemize}
 
 The action in Eq.~\eqref{eq: second order action} maintains general applicability throughout the early universe evolution. While the potential $V(\phi)$ can in principle take various phenomenological forms, general analysis is constrained without specific observational data. However, during the transition to the RD and MD eras—a phase known as \textit{preheating}~\cite{Kofman:1997yn,Kofman:1994rk}—the inflaton oscillates around the minimum of its potential. In this regime, the potential is universally approximated as quadratic:
 \begin{equation}
 V(\phi) \approx \frac{1}{2}m^2\phi^2 + \mathcal{O}(\phi^3).
 \label{eq:oscillating potential}
 \end{equation}
 This oscillatory behavior allows for an analytical treatment of the particle production mechanism while maintaining the generality of our framework.

 Next, we need to transform action \eqref{eq: second order action} into the Hamiltonian which will be denoted by the creation and annihilation operators. First, we need to define the conjugate momentum 
 \begin{equation}
\pi(\eta,\vec{x})\equiv \frac{\delta L}{\delta{v}'(\eta,\vec{x})} ={v}'-\frac{{a}'}{a}v.
 \label{eq: conjugate momentum}
 \end{equation}
 Then, we construct the Hamiltonian via $H=\int d^3xd\eta(\pi v'-\mathcal{L} )$, 
 \begin{equation}   
 H= \frac{1}{2}\int  d^3x d\eta \left [ {\pi}^2+(\partial _i v)^2+\frac{ {a}'}{a}(\pi v+v\pi) +a^{2}V_{,\phi\phi}v^{2}\right ].
 \label{hamilton of standard case}
 \end{equation}
 With the application of Fourier decomposition and Fourier transformation,
 \begin{equation}
 \hat{v} (\eta,\vec{x})=\int \frac{d^3k}{(2\pi)^{3/2}} \sqrt{\frac{1}{2k}}(\hat{c }_{\vec{k} }^{\dagger}v_{\vec {k}}^{\ast }(\eta )e^{-i\vec{k\cdot }\vec{x}}+ {\hat{c}_{\vec k}}v_{\vec k}e^{i\vec{k\cdot }\vec{x}}),
 \label{v}
 \end{equation}
 \begin{equation}
 \hat{\pi} (\eta,\vec{x})=i\int \frac{d^3k}{(2\pi)^{3/2}}\sqrt{\frac{k}{2}}(\hat{c }_{\vec{k}}^{\dagger}u_{\vec k}^{\ast }(\eta )e^{-i\vec{k\cdot }\vec{x}}-\hat{c}_{\vec k}u_{\vec k}e^{i\vec{k\cdot }\vec{x}}).
 \label{pi}
 \end{equation}
 As a result, we obtain the Hamiltonian in terms of creation and annihilation operators in momentum space, 
 \begin{equation}
 \begin{split}
 \hat{H}_{k}&= (\frac{a^{2}}{2k}V_{,\phi\phi}+k)(\hat c_{anc}^{\dagger }\hat c_{anc}+\hat c_{\vec{k}}\hat c^{\dagger }_{\vec{k}})+(\frac{a^{2}}{2k}V_{,\phi\phi}+i\frac{{a}'}{a})\hat c_{\vec{k}}^{\dagger }\hat c_{anc}^{\dagger }\\&+(\frac{a^{2}}{2k}V_{,\phi\phi}-i\frac{{a}'}{a})\hat c_{\vec{k}}\hat c_{anc}, 
 \end{split}
 \label{eq:hamilton}
 \end{equation}
 where $\hat{c}_{anc}$ corresponds to the $\hat{c}_{-\vec k}$ compared with our previous works \cite{Liu:2021nzx,Li:2023ekd,Li:2024kfm}. One observes that the Hamiltonian \eqref{eq:hamilton} consistently reduces to the Hamiltonian for the inflation era, 
 \begin{equation}
 \begin{split}
 \hat{H}_{k}= k(\hat c_{anc}^{\dagger }\hat c_{anc}+\hat c_{\vec{k}}\hat c^{\dagger }_{\vec{k}})+i\frac{{a}'}{a}\hat c_{\vec{k}}^{\dagger }\hat c_{anc}^{\dagger }-i\frac{{a}'}{a}\hat c_{\vec{k}}\hat c_{anc}. 
 \end{split}
 \end{equation}
 Our starting wave function is Eq. \eqref{eq:wave function of close syetem}, and it satisfies the Schr{\"o}dinger equation, 
 \begin{equation}\label{eq:5.2}
 i\frac{d}{d\eta} \left | \Psi  \right \rangle_{\phi} =\hat{H}_{k} \left | \Psi  \right \rangle_{\phi}. 
 \end{equation}
 By using the Hamiltonian in Eq. \eqref{eq:hamilton}, we can obtain the evolution equations of $T_{k}(\eta)$ and $\phi_{k}(\eta)$ 
 \begin{equation}
 \begin{split}
 \\& {T}'_{k}=-\frac{4T_k^{2}}{\omega}\sinh{(\frac{\omega}{2T_k})}[\frac{a{}'}{a} \cos (2\phi_{k})+\frac{a^{2}}{2k}V_{,\phi\phi}\sin{(2\phi_{k})}],  \\&{\phi}'_{k}=\frac{a^{2}}{2k}V_{,\phi\phi}+k+2\cosh{(\frac{\omega}{2T_k})}[\frac{a{}'}{a}\sin (2\phi_{k}) -\frac{a^{2}}{2k}V_{,\phi\phi}\cos{(2\phi_{k})} ].
 \end{split} 
 \label{eq:evolution of squeezed angle and temperature;RD and MD}
 \end{equation}
 Similarly, when we ignore the potential energy term, we obtain the evolution equation in the inflation period. 
 \begin{equation}
 \begin{split}
 \\& {T}'_{k}=-\frac{4T_k^{2}}{\omega}\sinh{(\frac{\omega}{2T_k})}\frac{a{}'}{a} \cos (2\phi_{k}), 
 \\&{\phi}'_{k}=k+2\cosh{(\frac{\omega}{2T_k})}\frac{a{}'}{a}\sin (2\phi_{k}).
 \end{split} 
 \label{eq:evolution of squeezed angle and temperature;inflation}
 \end{equation}
Using the conformal time \eqref{eq:conformal time and scale factor} as the evolutionary parameter, we can trace the evolution of the modes $T_k$ and $\phi_k$ across different eras.
 \begin{figure}
 	\centering
 	\includegraphics[width=0.99\linewidth]{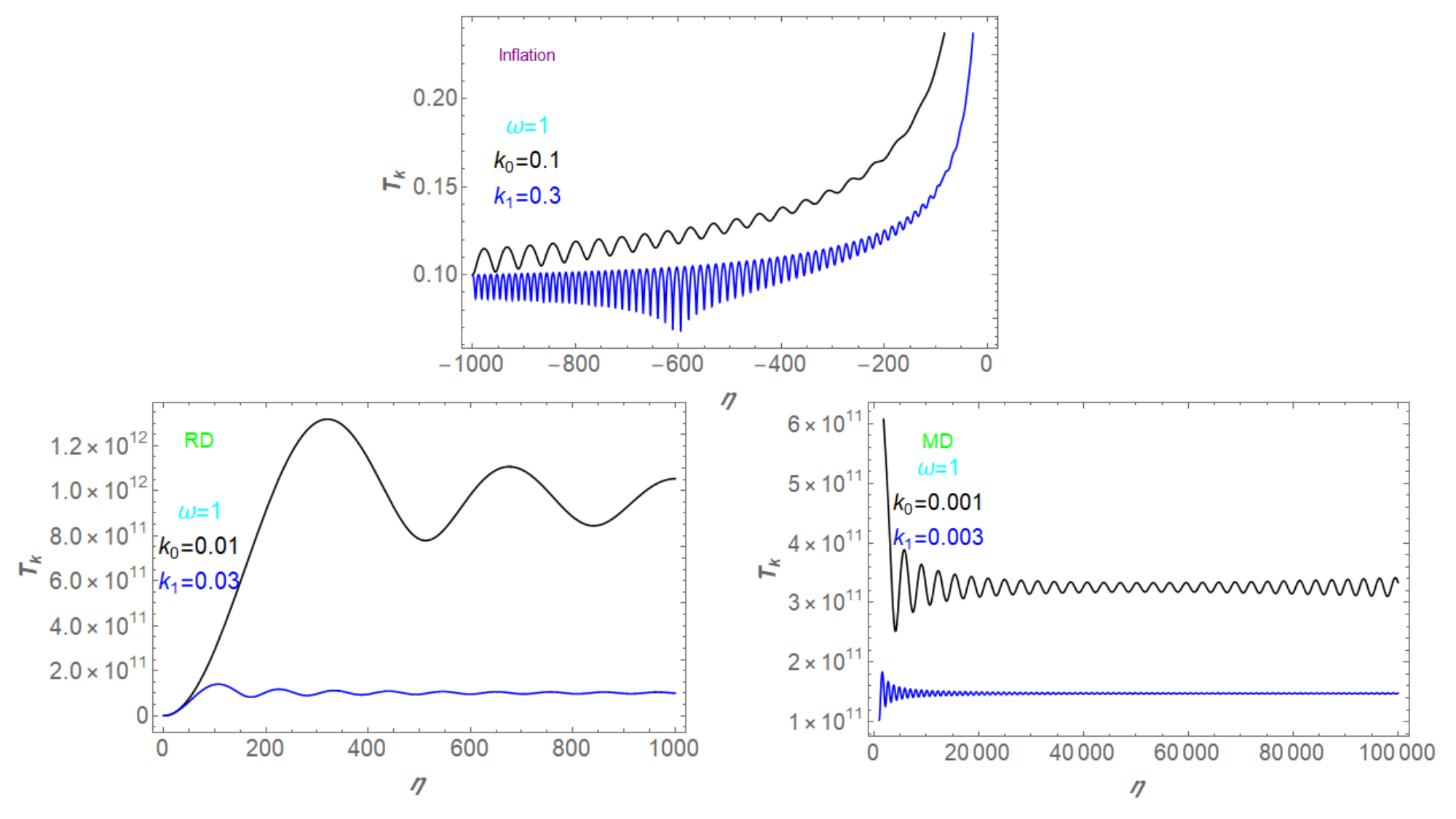}
 	\caption{The numerical solution for $T_{k}$ as a function of conformal time $\eta$ across three cosmological eras (inflation, RD, and MD) is presented. For computational simplicity, we adopt the parameter values: $\omega = 1$, $H_0 = 10^{-2.5}$, and $m = 10^{-6}$. The momentum scale $k$ assumes distinct characteristic values in each era, as illustrated in the respective panels.}
 	\label{fig:1}
 \end{figure}

 Figure~\ref{fig:1} illustrates the evolution of the effective temperature $T_k$ across the inflationary, RD, and MD eras. The numerical parameters are fixed at $\omega = 1$ (with the frequency contribution absorbed into the effective temperature), $m = 10^{-6}$, and $H_0 = 10^{-2.5}$.
 
 Each panel corresponds to a distinct wavenumber $k$. The figure reveals a sharp increase in effective temperature, particularly towards the end of inflation, signifying a rapid energy injection into the system during this epoch. In contrast, during the RD and MD eras, $T_k$ oscillates around a constant baseline. This behavior is directly attributable to the explicit inclusion of the inflationary potential.
 
 Physically, this saturation is linked to the preheating phase at the onset of the RD era---a highly non-equilibrium process characterized by particle production. Although preheating involves complex energy transfer from the inflaton to other species, the coherent oscillation of the inflaton field $\phi$ around its potential minimum ultimately drives the effective temperature toward a stable asymptotic value. Furthermore, the results demonstrate a clear momentum dependence: as shown in the respective panels, infrared modes (lower $k$) attain significantly higher effective temperatures.

 \begin{figure}
 	\centering
 	\includegraphics[width=1.0\linewidth]{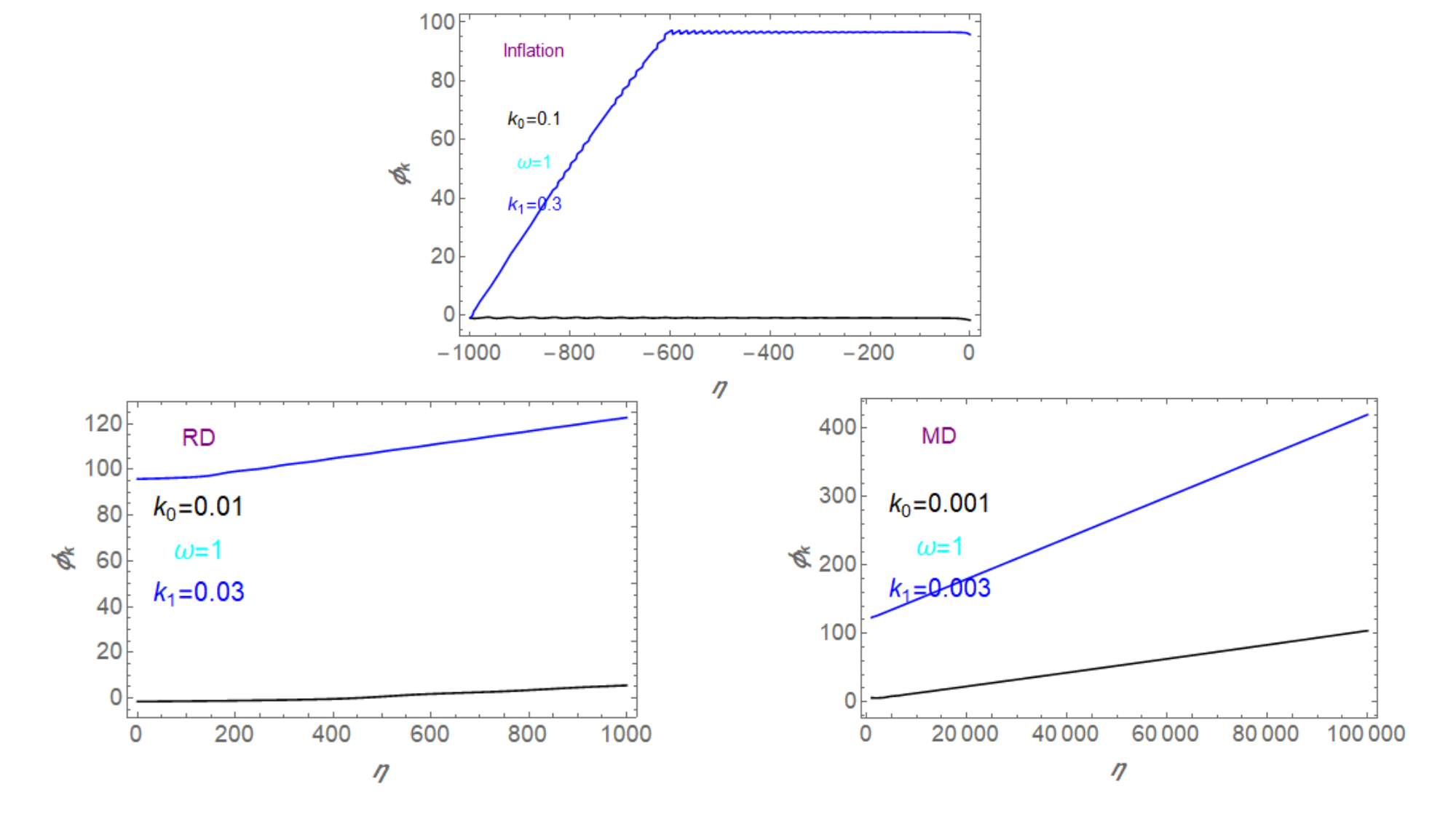}
 	\caption{The numerical solution for the squeezed angle $\phi_k$ as a function of conformal time $\eta$ is presented across three cosmological eras: inflation, RD era, and MD era. For simplicity, we adopt the parameter values $\omega = 1$, $H_0 = 10^{-2.5}$, and $m = 10^{-6}$. The momentum scale $k$ takes distinct characteristic values in each era, as illustrated in the respective panels.}
 	\label{fig:2}
 \end{figure}

 For completeness, we also present the numerical evolution of the squeezing phase $\phi_k$ across the inflationary, RD, and MD eras in Fig.~\ref{fig:2}. A distinct feature of the phase evolution---in contrast to the effective temperature $T_k$---is its momentum dependence: modes with larger wavenumbers $k$ exhibit more rapid growth throughout all epochs. Temporally, $\phi_k$ increases during the early stages of inflation before settling into a quasi-stationary plateau; notably, it resumes a growth trajectory upon transitioning into the subsequent RD and MD eras.
 
 As detailed in Sec.~\ref{sec:Evolution of temperature and squeezed angle}, our analysis departs from the standard Mukhanov-Sasaki formalism by utilizing the direct inflaton perturbation defined in Eq.~\eqref{eq: perturbation of inflaton}. This framework yields the second-order action (Eq.~\eqref{eq: second order action}), which explicitly retains the potential-dependent terms essential for capturing the dynamics beyond slow-roll. Solving the resulting evolution equations (Eq.~\eqref{eq:evolution of squeezed angle and temperature;RD and MD}) governs the dynamics of both parameters. Crucially, as will be demonstrated in the subsequent analysis, the Krylov complexity is determined \textit{exclusively} by the effective temperature $T_k$. Consequently, the saturation behavior of $T_k$ shown in Fig.~\ref{fig:1}---driven by the inclusion of the inflationary potential---serves as the primary dictator of the system's complexity evolution.

 \section{Krylov complexity with closed-system methodology}
 \label{sec: Krylov complexity with an approach of closed system}

 This section presents an investigation of the Krylov complexity and Krylov entropy (K-entropy), employing the numerical results for $T_k$ shown in Fig. \ref{fig:1}. Our analysis is based on the closed-system methodology detailed in \cite{Liu:2021nzx,Li:2021kfq,Li:2023ekd}.

 \subsection{Krylov complexity}
 \label{sec; Krylov complexity}
 
 To compute the Krylov complexity, we first express the closed-system wave function in terms of the Krylov basis expansion:
 \begin{equation}
 |\Psi\rangle_{\phi} = \sum_{n=0}^{\infty} i^n \phi_n(\eta) |\mathcal{O}_n) = \sqrt{1-e^{-\frac{\omega}{T}}} \sum^{\infty}_{n=0} e^{-2in\phi_{k}} e^{-\frac{n\omega}{2T}} |n\rangle \otimes |n\rangle_{\text{anc}}.
 \label{wave function for closed}
 \end{equation}
 We identify the reference state as the initial state, i.e., $|\mathcal{O}_0) \equiv |0\rangle \otimes |0\rangle_{\text{anc}}$. It is important to clarify that our derivation relies on the standard orthonormal inner product in the doubled Hilbert space, defined as $\langle m|n \rangle = \delta_{mn}$. Consequently, the inner product of the Krylov basis states satisfies $(\mathcal{O}_m|\mathcal{O}_n) = \delta_{mn}$, which is essential for the subsequent evaluations of Krylov complexity and for correctly recovering the Lanczos coefficients in Eq. \eqref{lanczos coefficient}.

 By comparing the coefficients of the Krylov expansion with the explicit TFD form, we extract the wave function amplitudes $\phi_n$:
 \begin{equation}
 \phi_{n} = \sqrt{1-e^{-\frac{\omega}{T}}} (-1)^{n} e^{-2in\phi_{k}} e^{-\frac{n\omega}{2T}}.
 \label{eq:Operator wave function}
 \end{equation}
 The dynamics of the system are governed by the Lanczos coefficients $b_n$, which appear in the tridiagonal action of the Liouvillian superoperator:
 \begin{equation}
 \mathcal{L}|\mathcal{O}_n) = b_{n+1}|\mathcal{O}_{n+1}) + b_n|\mathcal{O}_{n-1}).
 \label{resursion relation}
 \end{equation}
 Here, $b_n$ corresponds to the coefficients introduced in Section \ref{sec: Krylov complexity and Lanczos algorithm with an approach of closed system}. Treating the Liouvillian superoperator as an effective Hamiltonian expressed via creation and annihilation operators, and applying the specific total Hamiltonian from Eq.~\eqref{eq:hamilton} to the basis states $|\mathcal{O}_n)$, we derive the analytical expression for the Lanczos coefficients:
 \begin{equation}
 |b_n| = n\sqrt{\left(\frac{a^2 V_{,\phi\phi}}{2k}\right) + \left(\frac{a'^2}{a^2}\right)}.
 \label{lanczos coefficient}
 \end{equation}
 The positivity of the Lanczos coefficients is ensured by taking the absolute value. Finally, substituting the amplitudes $|\phi_n|^2$ into the definition of Krylov complexity, we obtain the explicit result:
 \begin{equation}
 \begin{split}
 K &= \sum_{n=0}^{\infty} n |\phi_{n}|^{2} \\
 &= \frac{e^{-\frac{\omega}{T}}}{1-e^{-\frac{\omega}{T}}} \\
 &= \frac{1}{2} e^{-\frac{\omega}{2T}} \sinh^{-1}{\left(\frac{\omega}{2T}\right)}.
 \end{split}
 \label{eq:krylov complexity of close syetem}
 \end{equation}

 \begin{figure}
 	\centering
 	\includegraphics[width=1.0\linewidth]{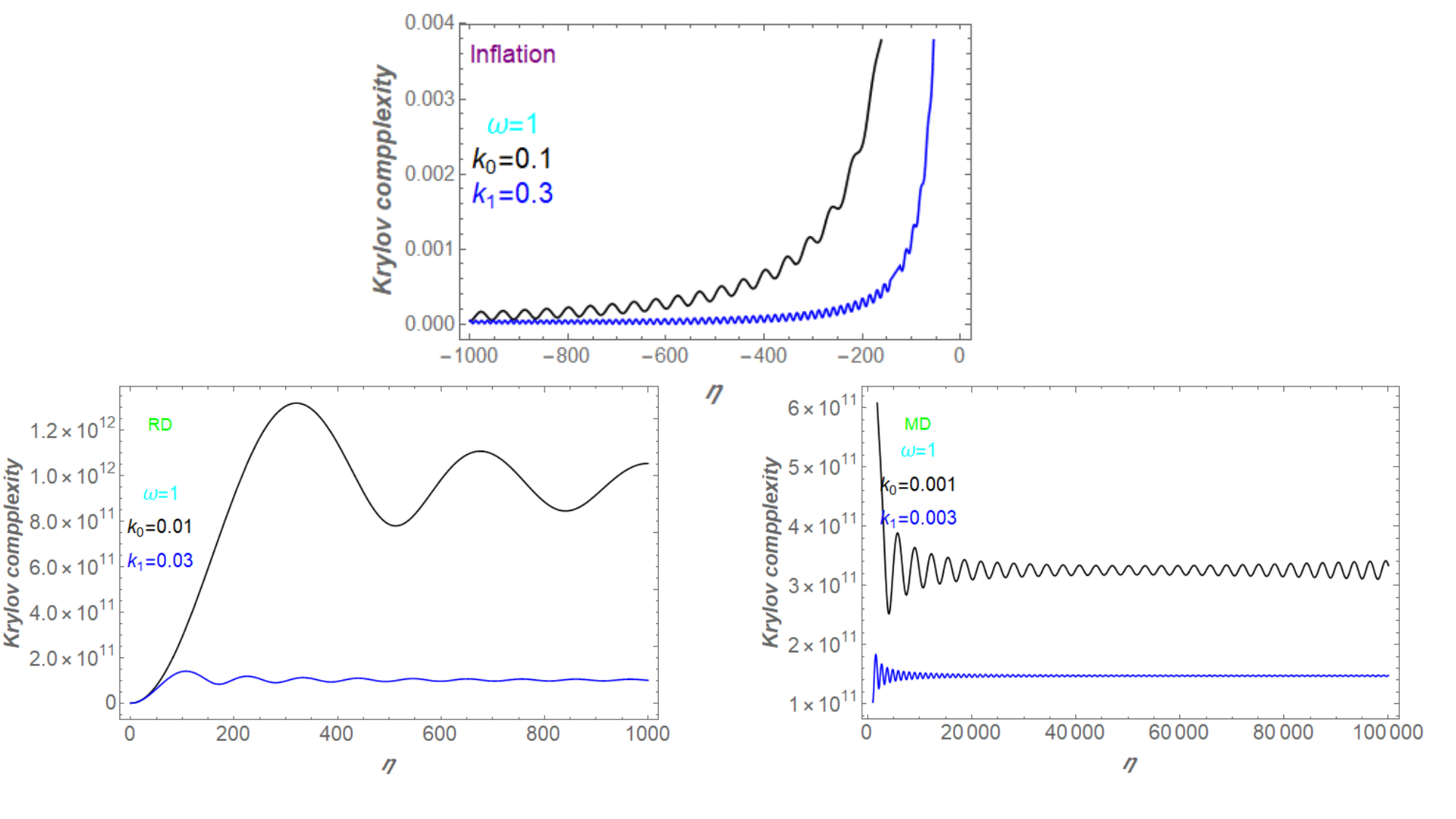}
 	\caption{The temporal evolution of Krylov complexity $C_K(\eta)$ is computed numerically for three cosmological phases: inflationary, RD, and MD. Parameter choices are $\omega = 1$, $H_0 = 10^{-2.5}$, and $m = 10^{-6}$, with era-specific momentum scales as shown in the corresponding panels.}
 	\label{fig:3}
 \end{figure}
 
The analytical expression for Krylov complexity explicitly demonstrates its independence from the squeezing phase $\phi_k$. The corresponding numerical evolution is presented in Fig.~\ref{fig:3}. Notably, the trajectory of Krylov complexity closely mirrors that of the effective temperature in Fig.~\ref{fig:1}, differing only by a scaling factor. During the inflationary epoch, the complexity exhibits monotonic growth, a result consistent with our previous findings~\cite{Li:2024kfm}. However, this behavior undergoes a distinct transition during the RD and MD phases: the inclusion of the potential leads to saturation and oscillations around a constant value, in agreement with the predictions of Ref.~\cite{Barbon:2019wsy}. This saturation is theoretically anticipated, as the RD and MD eras are characterized as weakly dissipative systems~\cite{Li:2024kfm}. We will further corroborate this phenomenon using open quantum system methodologies in the subsequent section.

 \subsection{Krylov Entropy }
 \label{krylov entropy}

 In quantum mechanics, entropy serves as a fundamental measure of information scrambling and system disorder. Within the Krylov complexity framework, this concept is quantified by the \textit{Krylov entropy} (or K-entropy). Here, we employ K-entropy to characterize the delocalization of curvature perturbations within the Krylov basis. Following Ref.~\cite{Barbon:2019wsy} and drawing an analogy to the von Neumann entropy, the K-entropy is defined as:
 \begin{equation}
 S_{K} \equiv -\sum^{\infty}_{n=0} |\phi_{n}|^{2} \ln |\phi_{n}|^{2}, 
 \label{eq:k_entropy_def}
 \end{equation}
 where the coefficients $\phi_n$ are provided in Eq.~\eqref{eq:Operator wave function}. By substituting the explicit expression for $\phi_n$ and performing the summation, we derive the analytical formula for the K-entropy:
 \begin{equation}
 \begin{split}
 S_{K} &= -\sum^{\infty}_{n=0} |\phi_{n}|^{2} \ln |\phi_{n}|^{2} \\
 &= \frac{\omega}{2T} e^{-\frac{\omega}{2T}} \sinh^{-1}{\left(\frac{\omega}{2T}\right)} - \ln{\left[2\sinh{\left(\frac{\omega}{2T}\right)}\right]} + \frac{\omega}{2T}.
 \end{split}
 \label{eq: k entropy}
 \end{equation}

 \begin{figure}
 	\centering
 	\includegraphics[width=1.01\linewidth]{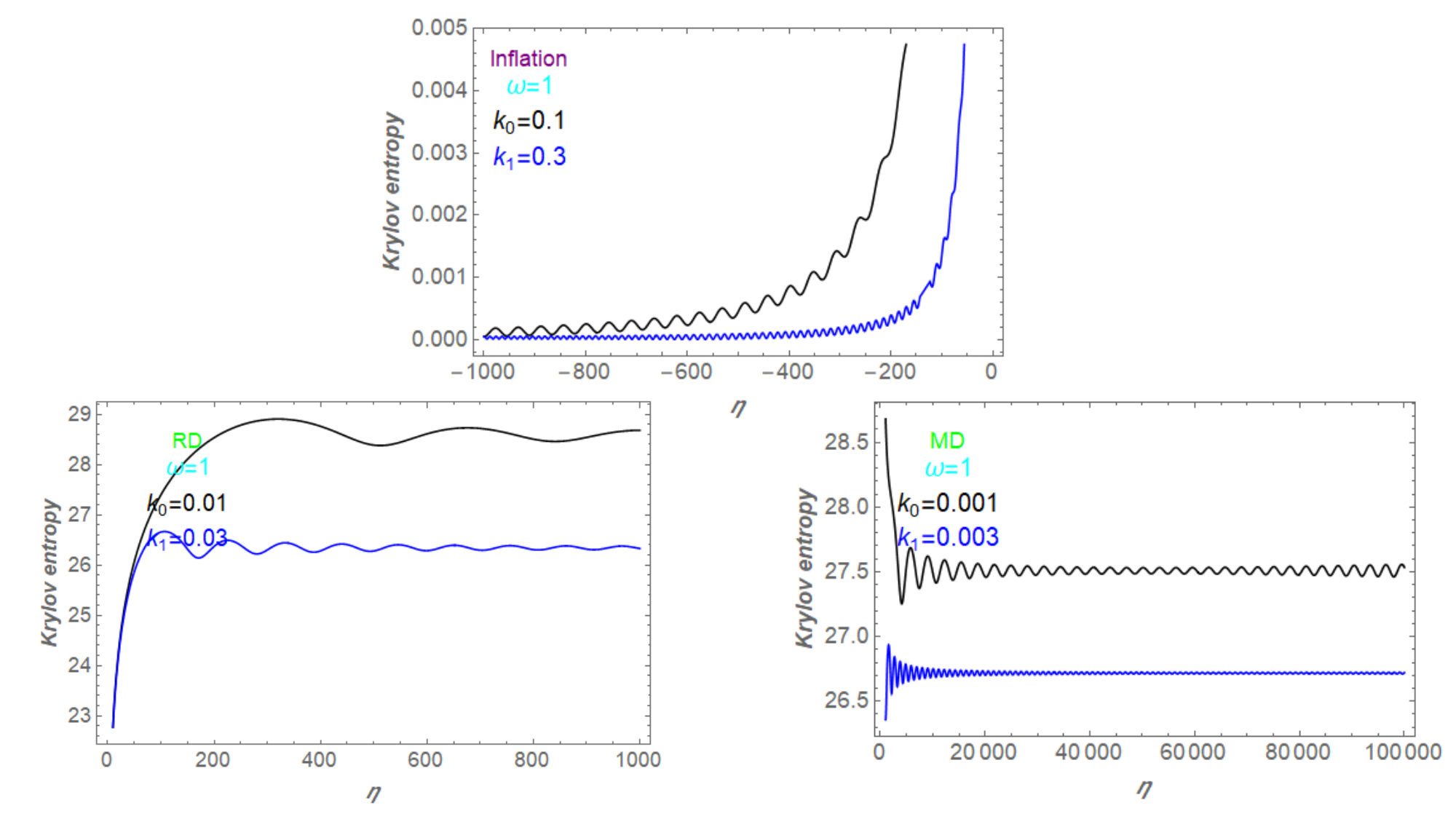}
 	\caption{We present the numerical evolution of the K-entropy with respect to $\eta$ during inflation, RD, and MD. For simplicity, the parameters are set as follows: $\omega=1$; $H_0=10^{-2.5}$; and $m=10^{-6}$. The momentum takes on different typical values in each of these distinctive periods.}
 	\label{fig:4}
 \end{figure}
The numerical evaluation of the K-entropy, based on Eq. \eqref{eq: k entropy}, is performed using the closed-system methodology.

Figure \ref{fig:4} illustrates that the K-entropy follows a trajectory nearly identical to that of the Krylov complexity shown in Fig. \ref{fig:3}. As entropy serves as a quantitative measure of disorder, its growth signifies that the system evolves toward a progressively chaotic state during inflation. This measure of chaos peaks at the onset of the RD era. Subsequently, extensive particle production via preheating drives the entropy to stabilize, oscillating around a constant asymptotic value.

Section \ref{sec: Krylov complexity with an approach of closed system} presents the comprehensive evolution of Krylov complexity, characterized by monotonic growth during inflation followed by saturation in the RD and MD eras. The K-entropy, which quantifies the degree of disorder in the universe, exhibits an analogous evolutionary pattern: the degree of chaos intensifies during inflation and eventually saturates. This implies that the system's disorder reaches its maximum capacity following the preheating phase (i.e., within the RD and MD regimes).

 \section{ Krylov complexity with open-system methodology}
 \label{sec:Krylov complexity with the approach of open system}
 
 As previously discussed, the RD and MD eras represent highly non-equilibrium systems due to prolific particle production. Furthermore, the loss of time-reversal invariance, evident from the non-conservation of energy under time reversal, which provides a compelling rationale for generalizing our approach to open-system methodology.

 \subsection{Lanczos coefficient and dissipative strength }
 \label{sec: Lanczos coefficient and dissipative strength }
Following the framework established in Refs.~\cite{Bhattacharya:2022gbz, Li:2024kfm}, we begin with the general expression for an operator in the Heisenberg picture:
\begin{equation}
\mathcal{O}(\eta) = e^{i\mathcal{L}_{o}\eta},
\end{equation}
where the Lindbladian $\mathcal{L}_{0}$, identified as the Hamiltonian in our model, possesses a well-defined action on the Krylov basis. From this action, we derive the recurrence relation first introduced in Ref.~\cite{Bhattacharya:2023zqt}:
\begin{equation}\label{eq:9.2}
\mathcal{L}_{o}|\mathcal{O}_{n}) = -ic_{n}|\mathcal{O}_{n}) + b_{n+1}|\mathcal{O}_{n+1}) + b_{n}|\mathcal{O}_{n-1}),
\end{equation}
where $c_n$ represents the diagonal component encoding the open-system information (dissipation), while $b_n$ denotes the standard Lanczos coefficient corresponding to the closed-system dynamics. While we interpret a non-vanishing $c_n$ as a signature of open-system dissipation in our specific framework, it is worth noting that $c_n \neq 0$ does not strictly necessitate non-unitary evolution. Depending on the choice of the inner product—for instance, if the associated correlation function is not an even function of time—non-zero $c_n$ contributions can also emerge in the Lanczos algorithm even when the system evolves unitarily (see, e.g., Sec. VIII A of \cite{Parker:2018yvk,Aguilar-Gutierrez:2025kmw}.

Based on this definition, the Lindbladian $\mathcal{L}_{o}$ for the open system is identified as the total Hamiltonian:
\begin{equation}
\mathcal{L}_{o} = \mathcal{H}_{o} = \mathcal{H}_{\text{close}} + \mathcal{H}_{\text{open}}.
\end{equation}
In our framework, the explicit form of the Hamiltonian corresponding to $\mathcal{L}_o$ is given by Eq.~\eqref{eq:hamilton}. Adopting the decomposition approach from Ref.~\cite{Li:2024kfm}, we identify the open-system component as
\begin{equation}\label{eq:open hamiltionian}
\mathcal{H}_{\text{open}} = \left(\frac{a^{2}}{2k}V_{,\phi\phi} + k\right)(\hat c_{\text{anc}}^{\dagger }\hat c_{\text{anc}} + \hat c_{\vec{k}}\hat c^{\dagger }_{\vec{k}}),
\end{equation}
and the closed-system component as
\begin{equation}\label{eq:close hamiltionian}
\mathcal{H}_{\text{close}} = \left(\frac{a^{2}}{2k}V_{,\phi\phi} + i\frac{a'}{a}\right)\hat c_{\vec{k}}^{\dagger }\hat c_{\text{anc}}^{\dagger } + \left(\frac{a^{2}}{2k}V_{,\phi\phi} - i\frac{a'}{a}\right)\hat c_{\vec{k}}\hat c_{\text{anc}},
\end{equation}
where the basis states are defined as $|\mathcal{O}_{n}) = |n\rangle \otimes |n\rangle_{\text{anc}}$. To verify this decomposition, we apply the Hamiltonian to the Krylov basis states. The action of the Hamiltonian on $|\mathcal{O}_n)$ yields the diagonal term $c_n = i(2n+1)\left( \frac{a^{2}}{2k} V_{,\phi\phi} + k \right)$. Consequently, the resulting expressions for $c_n$ and $b_n$ are:
\begin{equation}\label{eq:9.9}
c_{n} = i(2n+1)\left(\frac{a^{2}}{2k}V_{,\phi\phi} + k\right), \quad
|b_{n}| = n\sqrt{\left(\frac{a^{2}}{2k}V_{,\phi\phi}\right)^{2} + \left(\frac{a'}{a}\right)^{2}}.
\end{equation}
Crucially, the coefficients $c_{n}$ and $b_{n}$ in Eq.~\eqref{eq:9.9} are determined solely by the Hamiltonian structure and are independent of the effective temperature $T_k$. Specifically, $c_n$ encodes the dissipative characteristics, while $b_n$ quantifies the growth of operator complexity (or ``disorder'') within the system. Furthermore, matching our results with the general ansatz provided in Ref.~\cite{Bhattacharya:2022gbz}, we have:
\begin{equation}\label{eq:9.10}
b^{2}_{n} = |1-\mu^{2}_{1}|n(n-1+\beta), \quad c_{n} = i\mu_{2}(2n+\beta),
\end{equation}
where $\mu_1$ and $\mu_2$ are model parameters (with $\mu_1 = \mu_2$ corresponding to the single parameter $\mu$ in Ref.~\cite{Bhattacharya:2022gbz}), and $\mu_2$ serves as the dissipation coefficient. By comparing Eqs.~\eqref{eq:9.9} and \eqref{eq:9.10} with $\beta=1$, we derive the following correspondence:
\begin{equation} \label{eq:9.11}
|1-\mu_{1}|^{2} = \left(\frac{a^{2}}{2k}V_{,\phi\phi}\right)^{2} + \left(\frac{a'}{a}\right)^{2}, \quad
\mu_{2} = \left(\frac{a^{2}}{2k}V_{,\phi\phi} + k\right).
\end{equation}
It is evident that $\mu_1$ and $\mu_2$ are uniquely determined by the Hamiltonian in Eq.~\eqref{eq:hamilton}, offering a specific realization of the general framework in Ref.~\cite{Bhattacharya:2022gbz}.

We now discuss the behavior of $b_n$. As proposed by Parker et al. in Ref.~\cite{Parker:2018yvk}, the Lanczos coefficients for a chaotic many-body system in the thermal limit satisfy the asymptotic bound:
\begin{equation}
b_n \leq \alpha n + \eta,
\label{limit of bn}
\end{equation}
where $\alpha$ and $\eta$ characterize the system's chaotic properties. A system exhibits maximal chaos when this bound is saturated, i.e., $b_n = \alpha n + \eta$. In our analysis, we find $\eta = 0$, and the growth rate $\alpha$ is determined from Eq.~\eqref{eq:9.9} as:
\begin{equation}
\alpha = \sqrt{\left( \frac{a^{2}}{2k} V_{,\phi\phi} \right)^2 + \left( \frac{a'}{a} \right)^2 }.
\end{equation}
Since $b_n \propto n$, this result demonstrates that the early universe behaves as a maximally chaotic system in this regime. Furthermore, the Lyapunov exponent $\lambda_L$ is directly related to the growth rate $\alpha$ via:
\begin{equation}
\lambda_L = 2\alpha,
\label{lyapunov index}
\end{equation}
which is consistent with the relation $b_2 \approx 2\alpha = \lambda_L$ for large $n$ (or specifically $n=2$ in the linear ansatz). Since the Lanczos coefficients capture the chaotic dynamics, we utilize $b_n$ (and specifically $\alpha$) as a primary indicator of chaos in our subsequent analysis.

 \begin{figure}
 	\centering
 	\includegraphics[width=1.0\linewidth]{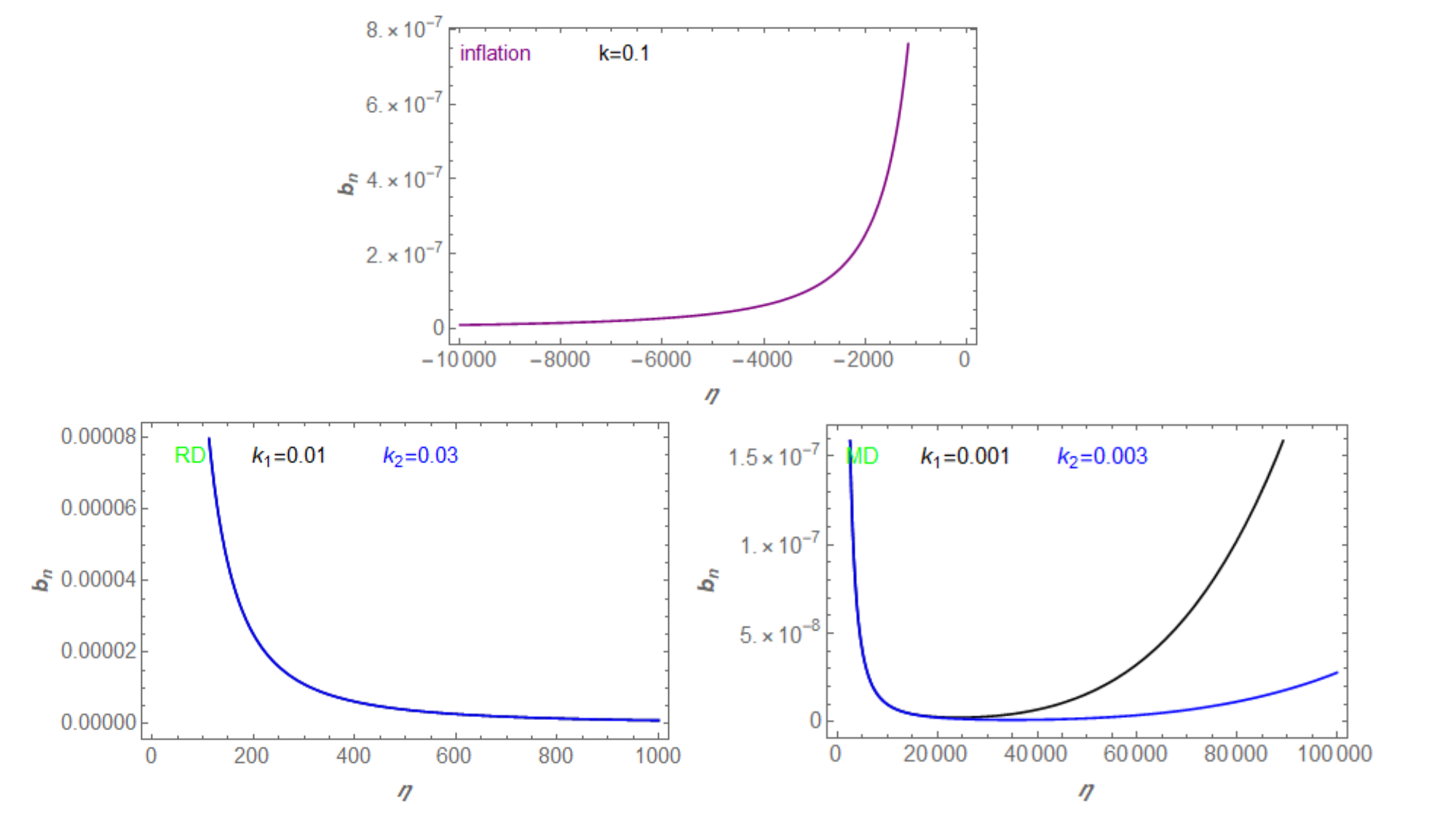}
 	\caption{Numerical results for the Lanczos coefficients $b_n(\eta)$ are shown for inflation, RD, and MD eras. We set $b_n=2\alpha=\lambda$ (at $n=2$), $H_0=10^{-2.5}$, and $m=10^{-6}$. The momentum scale $k$ is chosen appropriately for each era.}
 	\label{fig:5}
 \end{figure}
Figure~\ref{fig:5} illustrates the temporal evolution of the Lanczos coefficient $b_n$ as a function of conformal time $\eta$, spanning the inflationary, RD, and MD eras. The panels correspond to different momentum modes. In the first panel (inflation), $b_n$ exhibits monotonic growth, signaling an intensification of chaotic dynamics. In this regime, the coefficient follows the approximation $b_n \approx n \frac{a'}{a}$, which is independent of the wavenumber $k$ due to the suppression of potential contributions under slow-roll conditions. 
However, upon transitioning into the RD and MD eras, the overall chaotic intensity diminishes. While higher values of $k$ yield a slight enhancement in $b_n$, the order of magnitude remains comparable. This suppression is attributed to two primary factors: first, the breakdown of the slow-roll approximation, where potential terms become significant; and second, the distinct evolution of the scale factor $a(\eta)$ compared to the inflationary epoch. Physically, the preheating phase---characterized by extensive particle production---acts to dampen chaotic behavior during the MD era. Subsequently, we observe a modest resurgence in chaos at specific scales, driven by momentum amplification.

 \subsection{The wave function with open-system methodology}

 We adopt the open quantum system framework outlined in Ref.~\cite{Parker:2018yvk} to construct the system's wave function. As a prerequisite, we verify the validity of this method by analyzing the asymptotic behavior of the Lanczos coefficients $b_n^{(1)}$:
 \begin{equation}
 b_{n}^{(1)} = \alpha n + \gamma.
 \label{bn1}
 \end{equation}
 This linear growth signifies maximal chaos, a characteristic feature of infinite, non-integrable, many-body chaotic systems. This behavior is consistent with our specific calculation in Eq.~\eqref{eq:9.9}. Distinguishing our work from the original approach in Ref.~\cite{Parker:2018yvk}, we introduce two distinct parameters, $\mu_1$ and $\mu_2$ (defined in Eq.~\eqref{eq:9.10}), which are uniquely determined by the Hamiltonian via Eq.~\eqref{eq:9.11}. In the specific limit where $\mu_1 = \mu_2 = \mu$, our formulation reduces to the standard method.
 
 While Ref.~\cite{Parker:2018yvk} relied on a single parameter $\mu$ (typically constant, with $\mu \ll 1$ denoting weak dissipation), our generalized approach via Eq.~\eqref{eq:9.10} accommodates more realistic scenarios. This flexibility allows for the decomposition of various quantum inflationary models, including both single-field and multi-field scenarios. In the present work, based on the Hamiltonian in Eq.~\eqref{eq:hamilton}, we restrict our focus to single-field inflation. Leveraging this extended framework, Ref.~\cite{Li:2024kfm} derived the generalized wave function amplitudes:
 \begin{equation}
 \label{wave function for open}
 \phi_{n} = \frac{\operatorname{sech} \eta}{1+\mu_{2}\tanh \eta} |1-\mu^{2}_{1}|^{\frac{n}{2}} \left( \frac{\tanh \eta}{1+\mu_{2}\tanh \eta} \right)^{n}.
 \end{equation}
 In previous studies~\cite{Li:2024kfm}, we validated this approach by demonstrating that the Krylov complexity and K-entropy recover the closed-system results under the weak dissipation approximation. Furthermore, in Ref.~\cite{Zhai:2024odw} (Section 6.3), we explicitly showed that the weak dissipative limit of Eq.~\eqref{wave function for open} exactly reproduces the two-mode squeezed state (Eq.~\eqref{eq:two_mode_squeezed_state}) when identifying the parameter $\eta$ with the squeezing parameter $r_k$ (recalling the relation between $\beta\omega$ and $r_k$ in Eq.~\eqref{corresponding relation}). Building on these established relations, we construct the open-system wave function in terms of the effective temperature:
 \begin{equation}
 \phi_{n} = \frac{\sqrt{1-e^{-\frac{\omega}{T}}}}{1+\mu_{2}e^{-\frac{\omega}{2T}}} |1-\mu^{2}_{1}|^{\frac{n}{2}} \left( \frac{e^{-\frac{\omega}{2T}}}{1+\mu_{2}e^{-\frac{\omega}{2T}}} \right)^{n},
 \label{eq:wave function with T for open}
 \end{equation}
 where we have utilized the identifications $\operatorname{sech} \eta \leftrightarrow \sqrt{1-e^{-\frac{\omega}{T}}}$ and $\tanh \eta \leftrightarrow e^{-\frac{\omega}{2T}}$. Substituting these amplitudes into the general definition (Eq.~\eqref{eq:3.11}), we derive the explicit formula for Krylov complexity:
 \begin{equation}
 K \equiv \frac{|1-\mu_{1}^{2}|e^{-\frac{\omega}{T}}(1-e^{-\frac{\omega}{T}})}{\left[1+2\mu_{2}e^{-\frac{\omega}{2T}}+(\mu_{2}^{2}-|1-\mu_{1}^{2}|)e^{-\frac{\omega}{T}}\right]^{2}}.
 \label{eq:krylov complexity of open system}
 \end{equation}
 Using the correspondence $\beta\omega = -\ln\tanh^2 r_k$, we verify that this result aligns with the findings in Ref.~\cite{Li:2024kfm}. In the limit of weak dissipation ($\mu_1 \ll 1$ and $\mu_2 \ll 1$), the leading-order expansion of Eq.~\eqref{eq:krylov complexity of open system} yields:
 \begin{equation}
 K = \frac{1}{2}e^{-\frac{\omega}{2T}}\sinh^{-1}{\left(\frac{\omega}{2T}\right)} + \mathcal{O}(\mu_2).
 \label{leading order of krylov complexity}
 \end{equation}
 The leading term in Eq.~\eqref{leading order of krylov complexity} precisely matches the closed-system Krylov complexity derived in Eq.~\eqref{eq:krylov complexity of close syetem}, while $\mathcal{O}(\mu_2)$ represents the higher-order corrections arising from open-system effects. With the analytical expression established, we proceed to numerically simulate the evolution of Krylov complexity as a function of conformal time $\eta$.

 \begin{figure}
 	\centering
 	\includegraphics[width=0.95\linewidth]{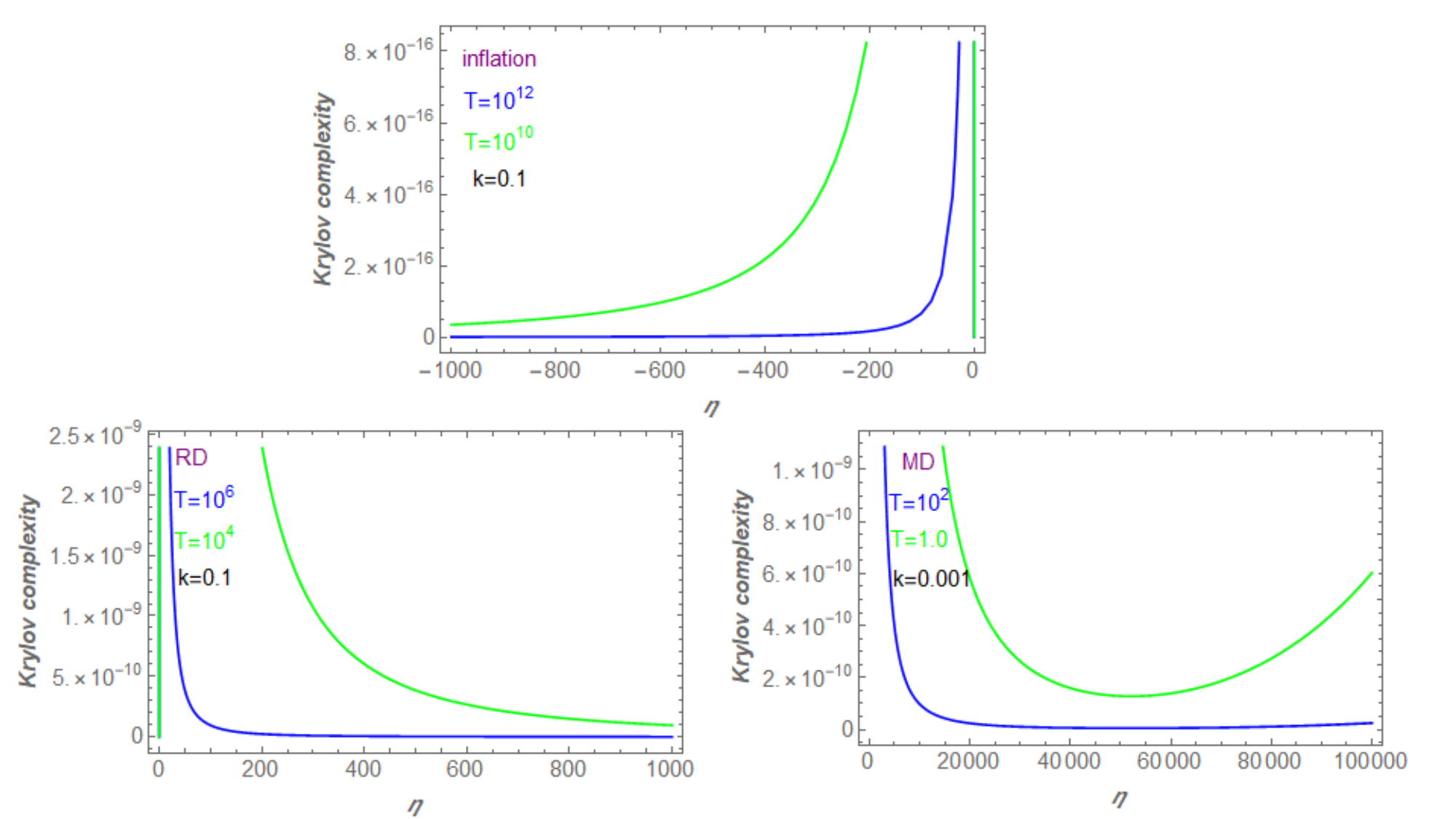}
 	\caption{We numerically evaluate the Krylov complexity from Eq.~\eqref{eq:krylov complexity of open system} as a function of conformal time $\eta$ across three cosmological eras: inflation, 
 		RD, and MD eras. For simplicity, we adopt the parameter values $\omega = 1$, $H_0 = 10^{-2.5}$, and $m = 10^{-6}$. The momentum and effective temperature parameters are assigned distinct values appropriate for each cosmological period.}
 	\label{fig:6}
 \end{figure}
 
Figure~\ref{fig:6} presents the numerical evolution of the Krylov complexity, computed via Eq.~\eqref{eq:krylov complexity of open system}, spanning the inflationary, RD, and MD eras. In each panel, the effective temperature $T$ and momentum $k$ are selected to reflect the characteristic energy scales of the respective epoch. We analyze the specific dynamics of each phase as follows:

\noindent\textbf{(a) Inflationary era:} The Krylov complexity exhibits a consistent growth trend throughout the inflationary epoch. Notably, an inverse relationship is observed: the rate of growth diminishes as the effective temperature increases.

\noindent\textbf{(b) RD era:} Upon transitioning into the RD era, the complexity follows a declining trajectory before saturating at a quasi-stationary value. This behavior aligns with the closed-system dynamics depicted in Fig.~\ref{fig:3}. The suppression is predominantly driven by the effective temperature rather than the momentum $k$. The convergence in evolutionary trends between the open and closed frameworks is physically attributed to the significant contributions from the inflaton potential.

\noindent\textbf{(c) MD era:} The dynamical behavior in the MD era largely mirrors that of the RD phase. However, a distinguishing feature emerges: for lower effective temperatures, the Krylov complexity exhibits a localized enhancement at specific scales. We identify this phenomenon as a direct consequence of particle production effects characteristic of this era.

 \subsection{K-entropy with open-system methodology}
 \label{sec: K-entropy with the approach of open system}
Having established the Krylov complexity using Eq.~\eqref{eq:wave function with T for open} and confirmed that its leading-order behavior recovers the closed-system case under the weak dissipative approximation, we now turn to the K-entropy. To provide a complete characterization of the system's dynamics, we calculate the K-entropy based on Eq.~\eqref{eq:wave function with T for open} and examine its temporal evolution. Substituting the wave function amplitudes into the general definition,
\begin{equation}
S_{K} = -\sum^{\infty}_{n=0} |\phi_{n}|^{2} \ln |\phi_{n}|^{2},
\end{equation}
we obtain the explicit expression:
\begin{equation}\label{eq:11.2}
\begin{split}
S_{K} =& -(1-e^{-\frac{\omega}{T}})\frac{(1+\mu_{2}e^{-\frac{\omega}{2T}})^{2}\left[\ln{(1-e^{-\frac{\omega}{2T}})} - 2\ln{(1+\mu_{2}e^{-\frac{\omega}{2T}})}\right]}{\left[1+2\mu_{2}e^{-\frac{\omega}{2T}}+(\mu_{2}^{2}-|1-\mu_{1}^{2}|)e^{-\frac{\omega}{T}}\right]^{2}} \\
& -(1-e^{-\frac{\omega}{T}})\frac{|1-\mu_{1}^{2}|e^{-\frac{\omega}{T}}\left[\ln{|1-\mu_{1}^{2}|} + e^{-\frac{\omega}{T}} - \ln{(1-e^{-\frac{\omega}{T}})}\right]}{\left[1+2\mu_{2}e^{-\frac{\omega}{2T}}+(\mu_{2}^{2}-|1-\mu_{1}^{2}|)e^{-\frac{\omega}{T}}\right]^{2}}.
\end{split}
\end{equation}
In the weak dissipation regime, characterized by $\mu_1 \ll 1$ and $\mu_2 \ll 1$, this expression reduces to the closed-system limit:
\begin{equation}
\begin{split}
S_{K} = \frac{\omega}{2T}e^{-\frac{\omega}{2T}}\sinh^{-1}{\left(\frac{\omega}{2T}\right)} - \ln{\left[2\sinh{\left(\frac{\omega}{2T}\right)}\right]} + \frac{\omega}{2T} + \mathcal{O}(\mu_2).
\end{split}
\end{equation}
This leading-order result is in exact agreement with Eq.~\eqref{eq: k entropy}. The fact that both the Krylov complexity and K-entropy successfully recover their closed-system counterparts confirms the validity of the open-system wave function constructed in Eq.~\eqref{eq:wave function with T for open}.
  \begin{figure}
 	\centering
 	\includegraphics[width=0.98\linewidth]{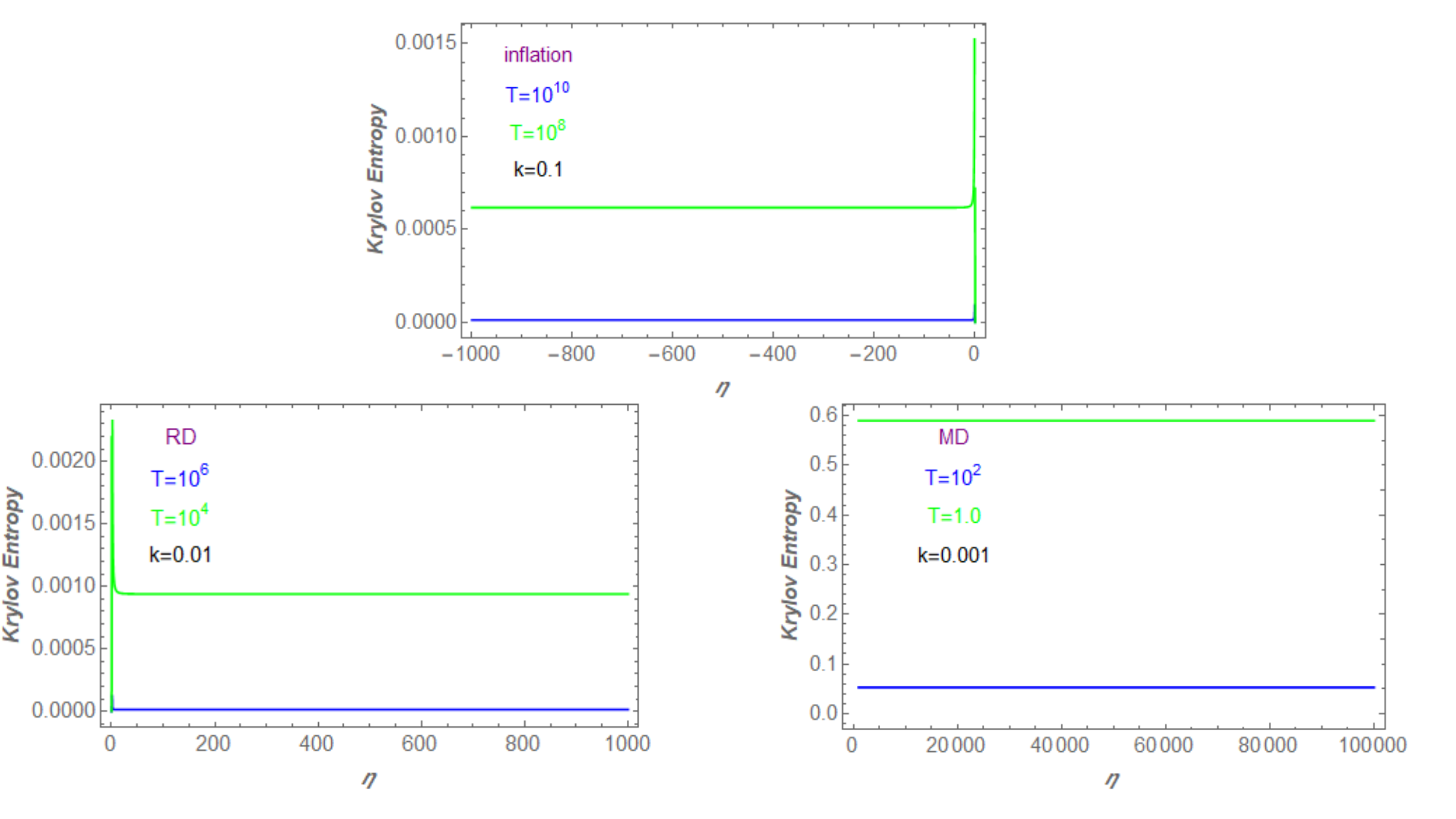}
 	\caption{We numerically evaluate the K-entropy from Eq.~\eqref{eq:11.2} as a function 
 		of conformal time $\eta$ across three cosmological eras: inflation, RD, and MD eras. For simplicity, we fix the parameters $\omega = 1$, $H_0 = 10^{-2.5}$, and $m= 10^{-6}$, while assigning appropriate values to momentum 
 		and effective temperature parameters for each distinct cosmological period.}
 	\label{fig:7}
 \end{figure}
 
Figure~\ref{fig:7} depicts the temporal evolution of K-entropy, calculated via Eq.~\eqref{eq:11.2}, throughout the early universe. The parameters adopted here are identical to those in Fig.~\ref{fig:6}. Interpreting these results from a thermodynamic perspective, we observe that the system's disorder increases monotonically during inflation, with the growth accelerating towards the conclusion of the epoch. Subsequently, the intensity of chaotic dynamics diminishes and stabilizes. During the preheating phase, driven by particle production, the measure of chaos saturates at distinct asymptotic levels. Notably, the evolutionary trajectory of K-entropy within this open-system framework closely mirrors the behavior observed in the closed-system methodology (Fig.~\ref{fig:4}).

 \subsection{Some discussions for $\mu_2$}
 
In our previous study~\cite{Li:2024kfm}, we established that the universe during inflation behaves as a strongly dissipative system, characterized by the condition $\mu_2 \gg 1$. We further conjectured that the saturation of Krylov complexity observed during the RD and MD eras suggests a transition to a weakly dissipative regime. In this section, we substantiate this hypothesis.

First, the evolutionary profiles of the Krylov complexity---analyzed via both closed and open system methodologies in Figs.~\ref{fig:4} and \ref{fig:6}---demonstrate that the complexity indeed saturates to nearly constant asymptotic values. This behavior aligns with the characteristics of the weak dissipation regime discussed in Ref.~\cite{Parker:2018yvk}. A more rigorous justification is provided by the explicit expression for the dissipation coefficient $\mu_2$:
\begin{equation}
\mu_2 = \left(\frac{a^{2}}{2k}V_{,\phi\phi} + k\right).
\end{equation}
Using the parameter values adopted in Figs.~\ref{fig:4} and \ref{fig:6}, we confirm that the condition $\mu_2 \ll 1$ holds throughout both the RD and MD eras. From a physical perspective, the preheating phase during the early MD era involves the transfer of energy from the inflaton potential to other fields, resulting in significant particle production. This process facilitates the environmental transition from a strongly dissipative regime to a weakly dissipative one. Consequently, our numerical results validate the conjecture that the universe in the RD and MD eras operates as a weakly dissipative system.

This section summarizes the application of the open system approach to Krylov complexity in the early universe, as detailed in Sec.~\ref{sec:Krylov complexity with the approach of open system}. By employing this framework, we have numerically reconstructed the full evolution of both Krylov complexity and K-entropy.

During inflation, the Krylov complexity exhibits sustained growth, a feature attributable to the strongly dissipative nature of this epoch. As the universe transitions into the RD and MD eras, the complexity saturates to stable values. This saturation reflects the emergence of a weakly dissipative character, a behavior primarily governed by the evolution of the inflationary potential.

Regarding the K-entropy, our results indicate that the measure of chaos intensifies during inflation before stabilizing at constant levels during the RD and MD eras. This stabilization is a direct consequence of particle production during preheating. Therefore, we conclude that particle production modifies the cosmic environment, acting as the catalyst for the transition from a strongly dissipative regime to a weakly dissipative one.

 \section{Summary and outlook}
 \label{Summary and outlook}

 The growing prominence of Krylov complexity in high-energy physics motivates its application to the early universe, a history encompassing the inflationary, RD, and MD eras, during which all known matter and energy densities originated. It is thus of particular interest to investigate the behavior of Krylov complexity across this entire evolutionary timeline. Moreover, since particle production during the RD and MD eras is a highly non-equilibrium process, thermal effects must be taken into account. In this context, adopting a two-mode thermal state offers a more reliable framework than a pure two-mode squeezed state. Furthermore, while the inflationary potential is often neglected under the slow-roll approximation during inflation, it becomes significant upon the transition to the RD and MD eras as slow-roll conditions break down. Understanding the evolution of Krylov complexity during these later stages is therefore essential for constructing a unified description of its dynamics throughout the early universe.
 
 We have conducted a systematic study of Krylov complexity in the early universe, employing both closed- and open-system methodologies. The principal results are summarized below:
 
 \paragraph*{(a) Theoretical Framework.}
 In Section~\ref{sec:Single-Component Universe}, we established the fundamental framework for describing the early universe, encompassing the inflation, RD, and MD eras via the scale factor evolution presented in Eq.~\eqref{eq:conformal time and scale factor}. Since the thermal state constitutes a mixed state without an explicit wave function representation, we defined its purification using the thermofield double (TFD) formalism in Eq.~\eqref{eq:generalized TFD}. This construction establishes a direct correspondence between the squeezing parameter $r_k$ and the effective temperature $T_k$, enabling the thermal state to be represented as a two-mode pure state, as shown in Eq.~\eqref{eq:wave function of close syetem}.
 
 A primary objective of this work was to investigate the influence of the inflationary potential on Krylov complexity. Consequently, we deliberately avoided employing the standard Mukhanov–Sasaki variable, which inherently mixes contributions from the potential. Instead, in Section~\ref{sec:Evolution of temperature and squeezed angle}, we perturbed the inflaton field directly and performed covariant quantization based on the second-order action given in Eq.~\eqref{eq: second order action}. The resulting Hamiltonian (Eq.~\eqref{eq:hamilton}) explicitly retains the contribution from the inflationary potential.
 
 \paragraph*{(b) Closed-System Dynamics.}
 Following the closed-system methodology, we numerically computed the evolution of the effective temperature $T_k$ and phase $\phi_k$ with respect to conformal time $\eta$ (Figs.~\ref{fig:1} and~\ref{fig:2}). The evolution of effective temperature is of particular significance, as both the Krylov complexity and K-entropy depend exclusively on $T_k$. Our results indicate that while the effective temperature increases during inflation, it oscillates around approximately constant values during the RD and MD eras.
 
 The corresponding behavior of the Krylov complexity, displayed in Fig.~\ref{fig:3}, closely follows the trend of the effective temperature. Physically, this saturation can be attributed to the particle production mechanism during preheating in the RD and MD eras. The energy for this process originates from oscillations around the minimum of the inflationary potential, approximated by the quadratic form in Eq.~\eqref{eq:oscillating potential}. When the inflationary potential is explicitly included, the Krylov complexity saturates to nearly constant values (Fig.~\ref{fig:3}), a direct consequence of preheating dynamics. A similar evolutionary trend is observed for the K-entropy (Fig.~\ref{fig:4}). Recalling that K-entropy quantifies chaos in a dynamical system, our results suggest that the measure of chaos is enhanced during inflation and subsequently stabilizes during the RD and MD eras.
 
 \paragraph*{(c) Open-System Dynamics and Dissipation.}
 In both methodologies, the Lanczos coefficient (Eq.~\eqref{eq:9.10}) serves as a key indicator of dynamical chaos. As shown in Eq.~\eqref{eq:9.9}, this coefficient is explicitly determined by the Hamiltonian, and its behavior aligns with that of the K-entropy (Fig.~\ref{fig:5}). Within the open-system methodology, we adopted the generalized construction of the thermal state (Eq.~\eqref{eq:wave function with T for open}) to compute both the Krylov complexity (Eq.~\eqref{eq:krylov complexity of open system}) and K-entropy (Eq.~\eqref{eq:11.2}). The leading-order behavior of these quantities is consistent with the closed-system case, validating our wave function construction.
 
 Numerical results (Figs.~\ref{fig:6} and \ref{fig:7}) reveal a consistent trend: both measures grow during inflation and approach constant values in the RD and MD eras. Crucially, our analysis characterizes the inflationary period as a strongly dissipative regime, whereas the RD and MD eras emerge as weakly dissipative regimes. This supports the conclusion that the transition from a strongly dissipative system to a weakly dissipative one drives the saturation of Krylov complexity and K-entropy, consistent with earlier discussions in Ref.~\cite{Li:2024kfm}.
 
 \bigskip
 
 In summary, our analysis provides a comprehensive dynamical picture of the early universe from the perspective of quantum information. We demonstrate that Krylov complexity grows continuously during inflation—a phase characterized by strong dissipation. The subsequent transition to the RD and MD eras marks a saturation of this complexity. An identical trend is observed for the K-entropy, confirming their fundamental link. Based on these findings, we propose several directions for future work:
 
 \begin{enumerate}
 	\item[(a)] Refs.~\cite{Kofman:1997yn,Kofman:1994rk,Cheung:2007st} argue that the universe can be viewed as an open system. Conventionally, such systems are often described by non-Hermitian Hamiltonians. In our framework, although the Hamiltonian in Eq.~\eqref{eq:hamilton} is Hermitian, the corresponding wave function in Eq.~\eqref{eq:wave function with T for open} evolves non-unitarily. This implies that the combined system—comprising the quadratic action and the wave function—effectively constitutes an open quantum system. Future work will explore the generality of this approach. To obtain an explicit non-Hermitian Hamiltonian, one could extend the action in Eq.~\eqref{eq: total action} beyond quadratic order (e.g., including cubic terms in the Mukhanov–Sasaki variable) or consider more general potentials beyond the quadratic approximation.
 	
 	\item[(b)] During inflation, the curvature perturbation undergoes a quantum-to-classical transition. It is therefore crucial to investigate how decoherence~\cite{Burgess:2022nwu} influences Krylov complexity in the post-inflationary epochs. While the present work focuses on single-field inflation, future studies could extend this analysis to multi-field inflation~\cite{Liu:2019xhn,Liu:2020zzv,Liu:2021rgq,Zhang:2022bde} and $f(R)$ gravity~\cite{Liu:2020zlr,Liu:2018htf,Liu:2018hno}. Such extensions would allow us to probe how the geometry of field space and the non-trivial structure of modified gravity theories imprint themselves on the evolution of complexity and entropy.
 \end{enumerate}

 \section*{Acknowledgements}
LH and TL are supported by NSFC grant NO. 12165009, Hunan Natural Science Foundation NO. 2023JJ30487 and NO. 2022JJ40340.

\appendix
\section{Inflation, RD and MD}
\label{sec: inflation, RD and MD}


This appendix outlines the foundational framework of the early universe, driven by the dynamical evolution of the scale factor $a(t)$. As detailed in Sec.~\ref{sec:Single-Component Universe}, the cosmic history unfolds through three distinct epochs: inflation, the RD era, and the MD era.

The evolution initiates with inflation, a period of exponential expansion following the Big Bang. During this phase, cosmological perturbations originate from quantum fluctuations, which then transition to a classical description upon horizon exit.

Subsequently, the universe enters the RD era. In this regime, the universe is populated by relativistic particles, and the energy density evolves as $\rho_{\text{rad}}(t) = \rho_{\text{rad},0} (a_0/a(t))^4$, where $\rho_{\text{rad},0}$ and $a_0$ represent the initial energy density and scale factor, respectively. As the universe expands and cools, relativistic particles lose energy, eventually becoming non-relativistic. This cooling process leads to the onset of the MD era, characterized by an energy density scaling of $\rho_{\text{mat}}(t) = \rho_{\text{mat},0} (a_0/a(t))^3$. The transition implies an equality epoch where the radiation and matter densities become comparable, defined by the condition $\rho_{\text{rad}}(t) = \rho_{\text{mat}}(t)$.

\end{document}